\documentclass[pra,twocolumn,groupedaddress,showpacs]{revtex4-1}
\usepackage{amsmath,amssymb,graphicx,comment} 
\usepackage{color}

\newcommand{\ket}[1]{| #1 \rangle}
\newcommand{\bra}[1]{\langle #1 |}

\newcommand{\brakets}[2]{\langle #1|#2\rangle}

\newcommand{\beq}{\begin{equation}}
\newcommand{\eeq}{\end{equation}}
\newcommand{\bea}{\begin{align}}
\newcommand{\eea}{\end{align}}

\begin{document}

\title{Physical Resources for Quantum-enhanced Phase Estimation}

\author{Jaspreet Sahota}\email{jsahota@physics.utoronto.ca}
\author{Nicol\'as Quesada}
\author{Daniel F. V. James}
\affiliation{Centre for Quantum Information and Quantum Control, Department of Physics, University of Toronto, 60 Saint George Street, Toronto, Ontario M5S 1A7, Canada}



\begin{abstract}
We study the role of quantum entanglement (particle entanglement and mode entanglement) in optical phase estimation by employing the first and second quantization formalisms of quantum mechanics. The quantum Fisher information (QFI) is expressed as a function of the first and second order optical coherence functions. The resulting form of the QFI elucidates the deriving metrological resources for quantum phase estimation: field intensity and photon detection correlations. In addition, our analysis confirms that mode entanglement is not required for quantum-enhanced interferometry, whereas particle entanglement is a necessary requirement.

\end{abstract}


\maketitle 

\section{Introduction}

The discovery of quantum mechanics has paved the way for an imminent technological revolution: Information processing tasks (IPT) that successfully leverage quintessential quantum mechanical resources (e.g. entanglement) can substantially outperform their classical counterparts. Notable examples include quantum computing \cite{nielsen2010quantum}, quantum teleportation \cite{PhysRevLett.70.1895}, quantum cryptography \cite{RevModPhys.74.145}, quantum-enhanced photodetector calibration \cite{Migdall:02}, quantum imaging \cite{doi:10.1080/09500340600589382}, and quantum illumination \cite{Lloyd12092008}. The successful application of entanglement to achieve these otherwise impossible IPTs has fostered the belief that entanglement is an essential ingredient in attaining a quantum advantage over classical IPTs.

In this paper, we study the physical resources responsible for quantum-enhanced parameter estimation. For this particular IPT, the goal is to estimate an unknown parameter which is sampled by appropriately prepared quantum probes. The performance of this task, typically quantified by the variance in the estimated parameter, can be immensely enhanced by employing non-classical probes. For instance, if the quantum probe is a $n$ qubit state belonging to the Hilbert space $\mathcal{H}^{\otimes n}$ (where $\mathcal{H}$ is a single qubit Hilbert space) then entanglement between the $n$ qubits is required to surpass the classical estimation precision limit (i.e. the shot-noise-limit) \cite{pezze09}. In addition, interferometric phase estimation below the shot-noise-limit has been achieved using entangled states \cite{Hudelist:2014dz, PhysRevLett.59.278, PhysRevLett.59.2153, Goda:2008yg}. Such states typically contain both mode entanglement and particle (i.e. photon) entanglement -- prompting us to ask: how does phase sensitivity depend on the different types of entanglements studied in quantum optics?

In the following, we elucidate the physical resources that drive optical interferometric performance by analyzing the \emph{quantum Fisher information} (QFI) from two naturally arising and distinct perspectives, resulting from the first and the second quantization formalisms of quantum mechanics. We deduce the following properties of the probe state to be conducive to quantum-enhanced phase estimation: (i) the mean photon number of the probe state, which is proportional to the energy of the electromagnetic field corresponding to the probe, and (ii) the detection correlations exhibited by the photons, as qualified by the Glauber coherence functions. Finally, we illustrate the dependence of these photon correlations on mode entanglement and particle entanglement in order to determine the role of each type of entanglement in quantum-enhanced phase estimation. 

\section{Background}

\begin{figure}[b]
\centering
\includegraphics[width=0.45 \textwidth]{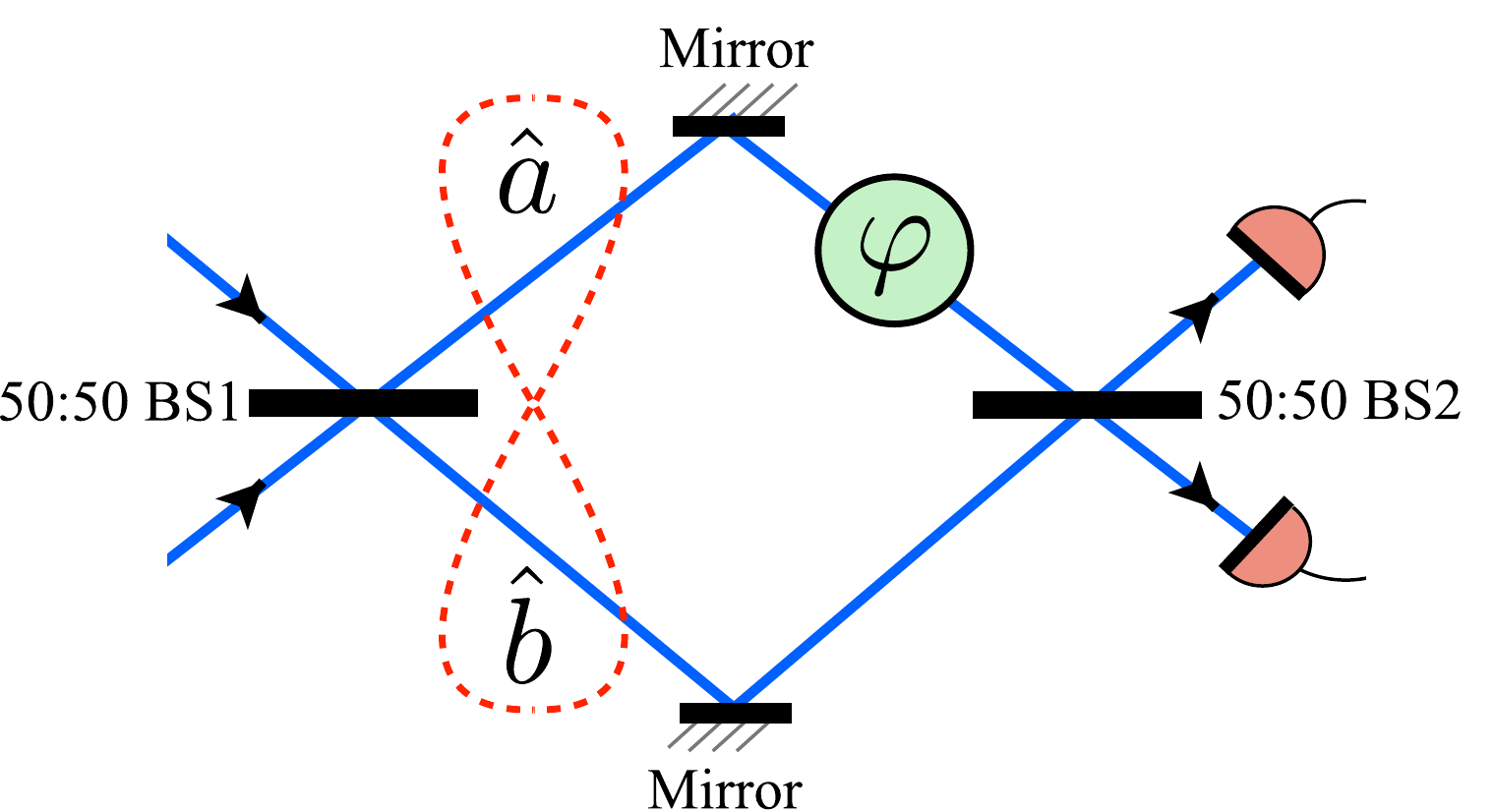}
\caption{(Color online)  Schematic diagram of the MZI. The arms inside the interferometer are represented by modes $\hat{a}$ and $\hat{b}$. The dotted red line indicated that the probe state, inside the interferometer, can potentially be mode entangled.}
\label{fig:MZI}
\end{figure} 

The role of mode entanglement has been studied for specific nonlinear quantum metrology systems by Tilma, \emph{et al.} \cite{PhysRevA.81.022108}, and independently by Datta and Shaji \cite{Datta2012}. Their work suggests that mode entanglement is not a vital resource for obtaining a quantum-advantage in nonlinear metrology. In the following we focus our discussion on the experimentally pertinent and readily accessible scenario of linear quantum metrology. In addition, instead of examining specific systems, as done in previous approaches, we provide a general argument about metrological resources using the QFI.

In linear quantum metrology, we are interested in measuring an unknown parameter $\varphi$ that has been encoded in a probe state $\ket{\Psi}$ via the unitary transformation $U = \exp{(i \varphi \hat{H})}$, generated by a linear Hamiltonian $\hat{H}$. If $\ket{\Psi} \in \mathcal{H}^{\otimes n}$, then entanglement between the $n$ qubits is required to overcome the shot-noise-limit \cite{pezze09}. Whereas, if $\ket{\Psi}$ does not describe a $n$ qubit probe, but instead a probe containing $n$ indistinguishable particles (e.g. $n$ photons), then the utility of entanglement for enhanced metrology is rather unclear. In the following, we study the role of quantum entanglement for this specific scenario.

\subsection{Quantum Metrology Tools}

\subsubsection{The Mach-Zehnder Interferometer and the Schwinger Representation} \label{MZISchwing}

The Mach-Zehnder interferometer (MZI) is the paradigm for studying linear quantum parameter estimation with optical systems. As illustrated in Fig. \ref{fig:MZI}, light is injected into a 50:50 beam-splitter (BS), resulting in a potentially mode entangled probe state inside the MZI. The bosonic creation and annihilation operators, $\hat{a}, \hat{a}^{\dagger}$ and $\hat{b}, \hat{b}^{\dagger}$, correspond to the two arms inside the MZI and thus describe different spatial modes of the probe state. The unknown parameter $\varphi$ is encoded into the probe state as a relative phase-shift between modes $\hat{a}$ and $\hat{b}$. After the phase-shift, the operation of the initial BS is reversed by using a second 50:50 BS. The light coming out of the final BS is measured \footnote{Measurements include: intensity measurements, homodyne, heterodyne, photon counting, parity measurements or a combination of these.}, and the resulting signal in processed in order to estimate $\varphi$. The operation of the MZI can be concisely described using the Schwinger representation of the angular momentum operators\cite{PhysRevA.33.4033}:
\begin{subequations}
\begin{align}
 \hat{J}_x &= \frac{1}{2}\left(\hat{a}^{\dagger} \hat{b} + \hat{b}^{\dagger} \hat{a} \right), \label{Schwinger1}\\
\hat{J}_y &= - \frac{i}{2}\left( \hat{a}^{\dagger} \hat{b} - \hat{b}^{\dagger} \hat{a} \right), \label{Schwinger2} \\
\hat{J}_z &= \frac{1}{2} \left( \hat{a}^{\dagger} \hat{a} - \hat{b}^{\dagger} \hat{b} \right), \label{Schwinger3} 
\end{align}
\end{subequations}
That is, the first and second BSs are described by $\exp{(-i \frac{\pi}{2} \hat{J}_x)}$ and $\exp{(i \frac{\pi}{2} \hat{J}_x)}$ respectively. The phase-shift is given by $\exp{(i \varphi \hat{J}_z)}$. If we denote the state injected into the first BS as $\ket{\Psi_o}$, then the probe state (the state immediately preceding the phase-shift operation) can be expressed as 
\beq
\ket{\Psi}=e^{-i \frac{\pi}{2} \hat{J}_x}\ket{\Psi_o}. \label{Probe}
\eeq 
The operators in Eq.(\ref{Schwinger1}) to Eq.(\ref{Schwinger3}) satisfy the SU(2) Lie algebra commutation relations
\begin{align}\label{su2}
[\hat J_k, \hat J_l ]=i \epsilon_{klm} \hat{J}_m,
\end{align}
where $\epsilon_{klm}$ is the Levi-Civita symbol,  and commute with the operator representing the total number of particles
\begin{align}
\hat n_a+\hat n_b= 2 \hat J_0.
\end{align}


\subsubsection{The Quantum Fisher Information} \label{sec:QFI}

Given expression (\ref{Probe}) for the MZI probe state, we can write the modified probe state (the state after the phase-shift has occurred) as $\ket{\Psi_\varphi}=\exp{(-i \varphi \hat{J}_z)}\ket{\Psi}$. Now we can quantify the phase-shift sensitivity of this particular probe by using the QFI, which for pure states is defined as
\beq
\mathfrak{F}=4 \left( \brakets{\Psi_\varphi^{\prime}}{\Psi_\varphi^{\prime}} - |\brakets{\Psi_\varphi^{\prime}}{\Psi_\varphi}|^2 \right), \label{QFI}
\eeq   
where the primes denote derivatives with respect to $\varphi$ \cite{Caves94}. The QFI is an indispensable tool in quantum metrology as it determines the ultimate lower-bound on the variance $\Delta \varphi$ of the estimated parameter attainable by using any locally unbiased estimator and by performing any quantum measurements on $\ket{\Psi_\varphi}$ \cite{Caves94, 1976quantum, holevo2011probabilistic}. This is expressed mathematically as the quantum Cram\'{e}r-Rao bound:
\beq
\Delta \varphi \ge \frac{1}{\sqrt{\mathfrak{F}}}. \label{QCRB}
\eeq   
Therefore, we will quantify the performance of a MZI phase estimation protocol using $\mathfrak{F}$ \footnote{Not always possible ADD REFERENCES}. In addition, we will discuss the physical properties of the probe state that can be manipulated in order to increase $\mathfrak{F}$. In the case of the MZI, $\mathfrak{F}= 4 \text{Var}[\hat{J}_z]$, where $\text{Var}[\hat{J}_z] = \bra{\Psi}\hat{J}^{2}_{z}\ket{\Psi} - \bra{\Psi}\hat{J}_{z}\ket{\Psi}^2$ is the variance in the phase-shift Hamiltonian. This follows directly from Eq.(\ref{QFI}) and the expression $\ket{\Psi_\varphi}=\exp{(-i \varphi \hat{J}_z)}\ket{\Psi}$ of the modified probe.

\subsection{Quantum Entanglement}

In order to facilitate the later discussion on the role of entanglement in phase estimation protocols, we review the definition of entanglement and how it manifests itself in different domains (i.e. mode vs. particle entanglements). 

Suppose the $n$-partite pure state $\ket{\Psi_n}$ belongs to the Hilbert space
\beq
 \mathcal{H}=\mathcal{H}_1\otimes\mathcal{H}_2\otimes \dots \otimes\mathcal{H}_n =\bigotimes_{i=1}^{n} \mathcal{H}_i. \label{fact}
\eeq  
The state $\ket{\Psi_n}$ is \emph{separable} if and only if   
\beq
 \ket{\Psi_n} =\bigotimes_{i=1}^{N} \ket{\psi^{(i)}},
\eeq  
where $\ket{\psi^{(i)}}$ can be any state in $\mathcal{H}_i$. We define $\ket{\Psi_n}$ to be entangled if and only if it is not separable. Therefore, it is evident that the entanglement of a quantum state depends on how we chose to partition its ambient Hilbert space. In quantum optics, there are two distinct and naturally arising ways of partitioning the Hilbert space: (i) we can partition with respect to the spatial modes of the electromagnetic field corresponding to the quantum state and (ii) we can partition with respect to the individual photons in the quantum state. In the following, we refer to approach (i) and approach (ii) as the mode picture and the particle picture respectively. 

\subsubsection{Mode Entanglement}

In the mode picture, the Hilbert space of the MZI probe state (\ref{Probe}) is partitioned with respect to the spatial modes corresponding to the two arms of the interferometer (Fig. \ref{fig:MZI}): $\mathcal{H}=\mathcal{H}_a\otimes\mathcal{H}_b$.  Let $\ket{\psi^{(a)}}$ be any state in $\mathcal{H}_a$. Likewise, let $\ket{\psi^{(b)}}$ be any state in $\mathcal{H}_b$. Then, by the definition given above, the probe state (\ref{Probe}) is mode entangled if and only if $\ket{\Psi} \ne \ket{\psi^{(a)}}\otimes\ket{\psi^{(b)}}$. In the mode picture, this system has a bipartite structure for which the notion of entanglement is succinctly quantified in terms of the \emph{Schmidt decomposition} \cite{nielsen2010quantum}.

\subsubsection{Particle  Entanglement} \label{particleEnt}

In the particle picture, we can define photon entanglement provided that the probe state contains precisely $n$ photons (i.e. there are no particle number fluctuations). We introduce pseudo-labels for these photons: $1, 2, \dots, n$, which can be used to partition the Hilbert space as in Eq. (\ref{fact}). Hence, we can define particle entanglement using the criterion of separability outlined above. Note that since the photons in the probe state are \emph{indistinguishable}, the labels we have introduced are fictitious and serve merely as a book-keeping tool. In other words, we are defining entanglement between operationally inaccessible sub-systems, which has provoked considerable debate in the past concerning the utility of such entanglement \cite{PhysRevA.67.013607, PhysRevA.67.013609, PhysRevA.70.012109, PhysRevLett.95.120502, PhysRevB.76.113304}. This issue was recently settled by Killoran, \emph{et al.} by constructing a protocol for converting identical-particle entanglement into entanglement between distinguishable spatial modes, which can be used to execute the various IPTs discussed above \cite{killoran14}. In section \ref{particlepic}, we will outline the type of photon entanglement required for quantum-enhanced MZI phase estimation.

\subsection{Optical Coherence of a MZI Probe} \label{coherences}

\begin{figure}[b]
\centering
\includegraphics[width=0.45 \textwidth]{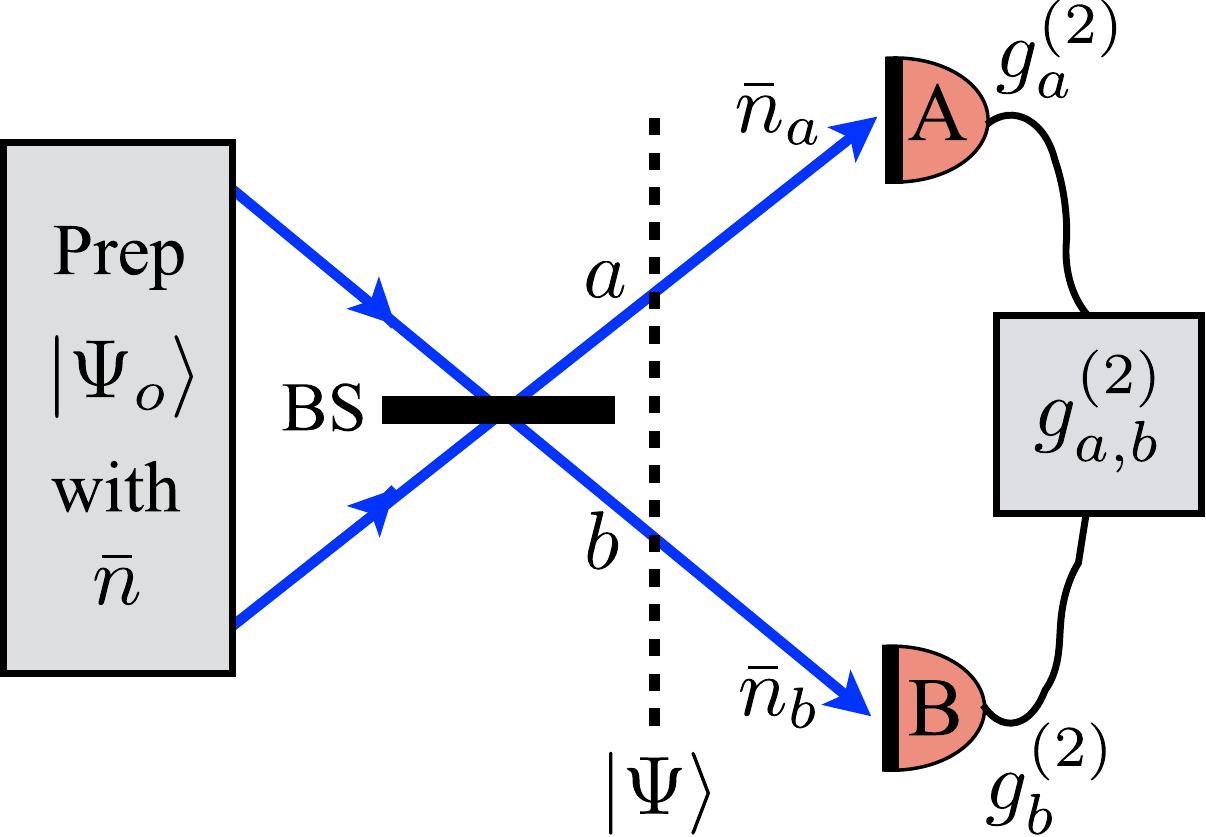}
\caption{(Color online)  A state $\ket{\Psi_o}$, with mean photon number $\bar{n}$, is injected to a 50:50 BS to create the probe state (\ref{Probe}). Photon detectors A and B intercept modes $a$ and $b$ of the probe $\ket{\Psi}$, respectively. The photon detection correlations, as quantified by the coherence functions: $g_a^{(2)}$, $g_b^{(2)}$, $g_{a,b}^{(2)}$, and the mean energy of the probe (which is proportional to $\bar{n}$) determine metrological performance and thus these parameters can be thought of as quantifiable resources for optical phase estimation.} 
\label{fig:probecorr}
\end{figure}

The quantum resources for optical phase estimation can be quantified using the concept of optical coherence. Therefore, we will review some relevant topics form optical conference theory in this section \cite{PhysRev.130.2529, PhysRevLett.10.277}.

An electric field
\begin{equation}
\hat{E}\left(\mathbf{r},t\right) = \hat{E}^{(+)}\left(\mathbf{r},t\right) + \hat{E}^{(-)}\left(\mathbf{r},t\right)
\end{equation}  
can be expanded, in a volume $V$, in terms of orthonormal spatial-temporal modes, such that
\beq
\hat{E}^{(+)} \left(\mathbf{r},t\right)=i \sqrt{\frac{\hbar \Omega}{2 \varepsilon_0 V}}  \sum_{k} \hat{a}_k u_k\left(\mathbf{r} \right)e^{i \Omega_k t}, \label{EPos}
\eeq 
and $\hat{E}^{(-)} =\hat{E}^{(+) \dagger}$ \cite{gerry05} \footnote{In general, the $\sqrt{\hbar \Omega}$ term in (\ref{EPos}) should be frequency dependent: $\sqrt{\hbar \Omega_k}$. However, in the optical regime the frequency profile corresponding to the states is highly peaked around a central frequency $\Omega$, hence  $\sqrt{\hbar \Omega_k}$ is readily approximated as $\sqrt{\hbar \Omega}$.}. The frequency and spatial profile of the $k^{\text{th}}$ mode are denoted as $\Omega_k$ and $u_k\left(\mathbf{r} \right)$, receptively. The bosonic creation and annihilation operators satisfy the commutation relations $[\hat{a}_k, \hat{a}^{\dagger}_l]=\delta_{k,l} \hat{\mathbf{1}}$. The electric field for the MZI probe (\ref{Probe}) has support on only two spatial modes, $a$ and $b$ (Fig. \ref{fig:probecorr}), defined as
\begin{align} 
\hat{a} \left(t\right) u_a \left(\mathbf{r}\right) &\equiv \sum_{k\in \mathcal{A}} \hat{a}_k u_k\left(\mathbf{r} \right)e^{i \Omega_k t} \label{a}  \\
\hat{b} \left(t\right) u_b \left(\mathbf{r}\right) &\equiv \sum_{k\in \mathcal{B}} \hat{a}_k u_k\left(\mathbf{r} \right)e^{i \Omega_k t} 
\end{align} 
where set $\mathcal{A}$ ($\mathcal{B}$) enumerates the modes that occupy the first (second) arm of the MZI (Fig. \ref{fig:probecorr}). Therefore, the positive frequency part of the MZI field is
\beq
\hat{E}^{(+)}_{\text{MZI}} \left(\mathbf{r},t\right)=i \sqrt{\frac{\hbar \Omega}{2 \varepsilon_0 V}}  \left[\hat{a}(t) u_a\left(\mathbf{r} \right)+\hat{b}(t) u_b\left(\mathbf{r} \right)\right], \label{EMZI}
\eeq  
which gives $\hat{E}_{\text{MZI}} =\hat{E}^{(+)}_{\text{MZI}}+\hat{E}^{(-)}_{\text{MZI}}$.

\subsubsection{Field Intensity and Photon Detection} 

Figure \ref{fig:probecorr} depicts two detectors, A and B, which intercept the fields in mode $a$ and $b$, respectively. What is the probability that one of the detectors will register a detection event? As outlined below, this probability is directly proportional to the intensity of the field incident on the detector in question -- which is also a resource for phase estimation (as discussed later in section \ref{modepic}).

A detection event is registered at detector A when a photon from mode $a$ (\ref{a}) is annihilated and absorbed by the detector, resulting in the production of a photoelectron, which is amplified to a macroscopically observable photocurrent. Therefore, the probe state $\ket{\Psi}$ is mapped to $\ket{f}\sim \hat{a} \left(t\right) \ket{\Psi}$. The transition probability for this process can be obtained using Fermi's golden rule \cite{gerry05}:   
\beq
\mathcal{P}_{\ket{\Psi} \to \ket{f}} \sim |\bra{f}\hat{a} \left(t\right) \ket{\Psi}|^2.
\eeq
Summing over the set of all possible final states $\{\ket{f}\}$ to which the probe can transition gives the probability of observing a detection at A. 
\begin{align} 
\mathcal{P}_{\text{detection}} &\sim \sum_{k}  |\bra{f}\hat{a}\left(t\right)\ket{\Psi}|^2 \nonumber \\
&= \bra{\Psi}\hat{a}^{\dagger}\left(t\right) \left(\sum_{f}\ket{f}\bra{f}\right)\hat{a} \left(t\right)\ket{\Psi} \nonumber \\
&= \bra{\Psi}\hat{n}_a \ket{\Psi}=\bar{n}_a,
\end{align}  
where $\hat{n}_a =\hat{a}^{\dagger}\left(t\right)\hat{a}\left(t\right)$. Likewise, we obtain that the probability of observing a detection at detector B is directly proportional to the intensity of the incident field, $\bar{n}_b$, on detector B. The total intensity $\bar{n}=\bar{n}_a+\bar{n}_b$ is conserved by the MZI operations (i.e. beamsplitters and phase-shifters). Hence, $\bar{n}$ is typically regarded as a metrology resource and important phase sensitivity limits are defined in terms of how the QFI scales with respect to $\bar{n}$. For example, $\mathfrak{F}\sim\bar{n}$ means shot-noise scaling and $\mathfrak{F}\sim\bar{n}^2$ means Heisenberg scaling. In section \ref{modepic}, we will express the QFI in terms of the filed intensities, $\bar{n}_a$ and $\bar{n}_b$, and the second order coherence functions for the MZI probe.

\subsubsection{Photon Detection Correlations}

The probability of detecting two photons simultaneously is proportional to the second order coherence function. Two photons are detected simultaneously at detector A provided that two photons are annihilated simultaneously in mode $a$, such that $\ket{\Psi} \to \ket{f} \sim \hat{a}^2\left(t\right)\ket{\Psi}$. The probability for such a transition occurring is proportional to $|\bra{f}\hat{a}^2 \left(t\right) \ket{\Psi}|^2$ \cite{gerry05}. Again, we sum over the set of all possible final states $\{\ket{f}\}$ to obtain the probability of observing a two photon simultaneous detection events at A. This gives
\beq
\mathcal{P}^{(2)}_{\text{detection}} \sim \bra{\Psi}\hat{a}^{\dagger 2} \hat{a}^{2}\ket{\Psi}, \label{prob2}
\eeq
where we have dropped the time dependence to simplify notation. Rescaling (\ref{prob2}), by dividing the right-hand-side by $\bar{n}^2_a$, gives the second order coherence function for mode $a$. We can define an analogous coherence functions for mode $b$ and simultaneously for both modes $a$ and $b$. Therefore, the second order MZI coherence functions are
\begin{align} 
g^{(2)}_{a} &\equiv \frac{\bra{\Psi}\hat{a}^{\dagger 2} \hat{a}^{2}\ket{\Psi}}{\bar{n}^2_a}, \label{ga} \\
g^{(2)}_{b} &\equiv \frac{\bra{\Psi}\hat{b}^{\dagger 2} \hat{b}^{2}\ket{\Psi}}{\bar{n}^2_b}, \label{gb} \\
g^{(2)}_{a,b} &\equiv \frac{\bra{\Psi}\hat{b}^{\dagger} \hat{a}^{\dagger} \hat{a} \, \hat{b} \ket{\Psi}}{\bar{n}_a \bar{n}_b}. \label{gab} 
\end{align}  
The function $g^{(2)}_{b}$ is proportional to the probability of observing two photons simultaneously at detector B. Whereas, $g^{(2)}_{a,b}$ describes the probability of observing one photon at detector A and another photon at detector B simultaneously. 

\section{The Mode Picture: the QFI as a function of Optical Coherence} \label{modepic}

As described in section \ref{sec:QFI}, the performance of a MZI phase estimation protocol can be quantified by the QFI $\mathfrak{F}= 4 \text{Var}[\hat{J}_z]$. Now we will analyze this expression in the mode picture, where $\hat{J}_z$ is defined in  Eq. (\ref{Schwinger3}). Expanding $\text{Var}[\hat{J}_z]$ gives
\beq
\mathfrak{F}=\text{Var}[\hat{n}_a] + \text{Var} [\hat{n}_b] - 2 \, \text{Cov}[\hat{n}_a, \hat{n}_b], \label{exQFI}
\eeq
where $\hat{n}_a=\hat{a}^{\dagger}\hat{a}$, $\hat{n}_b=\hat{b}^{\dagger}\hat{b}$, and $\text{Cov}[\hat{n}_a, \hat{n}_b]=\langle \hat{n}_a\otimes \hat{n}_b \rangle- \langle \hat{n}_a \rangle \langle \hat{n}_b \rangle$ is the covariance of $\hat{n}_a$ and $\hat{n}_b$. We infer from (\ref{exQFI}) that the phase sensitivity depends on intra-mode and inter-mode properties of the probe state, which are quantified by the variance of the photon number operator of each interferometric mode and the $\text{Cov}[\hat{n}_a, \hat{n}_b]$, respectively. This inter-mode and intra-mode correlation dependence was highlighted in Ref.\cite{sahota15} by expressing (\ref{exQFI}) in terms of the Mandel-$Q$ parameter and the Pearson correlation parameters for the two MZI modes.

Here we will express (\ref{exQFI}) as a function of first order coherence, namely $\bar{n}_a$ and $\bar{n}_b$, and the second order MZI coherence functions (\ref{ga}) -- (\ref{gab}) in order to highlight explicitly the dependence of the QFI on intra-mode and inter-mode correlations. We start by noting that, the commutation relation $[\hat{a},\hat{a}^{\dagger}]=1$ can be used to rewrite (\ref{ga}) as $g^{(2)}_{a} =1 + \text{Var}[\hat{n}_a]/\bar{n}_a^2   - 1/\bar{n}_a$. Therefore,
\beq
\text{Var}[\hat{n}_a] = \bar{n}_a + \bar{n}_a^2 \left(g^{(2)}_a-1\right). \label{vara}
\eeq    
Likewise, for mode $b$ we obtain  
\beq
\text{Var}[\hat{n}_b] = \bar{n}_b+ \bar{n}_b^2 \left(g^{(2)}_b-1\right). \label{varb}
\eeq    
Now using Eq.(\ref{vara}), Eq.(\ref{varb}), and the fact that $\text{Cov}[\hat{n}_a, \hat{n}_b] =  \bar{n}_a \bar{n}_b (g_{a,b}^{(2)} -1)$, we can write Eq.(\ref{exQFI}) as
\begin{align}
\mathfrak{F} =& \, \bar{n} + \bar{n}_a^2 \left(g^{(2)}_a-1\right) + \bar{n}_b^2 \left(g^{(2)}_b-1\right) \label{Fmode}  \\
&-2 \bar{n}_a \bar{n}_b \left(g^{(2)}_{a,b}-1\right). \nonumber
\end{align}
This equation expresses the optimal MZI phase sensitivity as a function of physically accessible properties of the probe (cf. Fig. \ref{fig:probecorr}), namely the first and second order optical coherences: the field intensities and the second order Glauber coherences functions, respectively. Eq. (\ref{Fmode}) takes a particularly concise form when the probe $\ket{\Psi}$ is taken to be a path-symmetric state: i.e. $\ket{\Psi}$ is invariant with respect to an exchange (nonphysical) of modes $a$ and $b$. For such states $\bar{n}_a=\bar{n}_b=\bar{n}/2$ and $g^{(2)} \equiv g^{(2)}_a=g^{(2)}_b$. Therefore, (\ref{Fmode}) becomes
\beq
\mathfrak{F} =\bar{n} + \frac{\bar{n}^2}{2} \left(g^{(2)}-g^{(2)}_{a,b}\right). \label{ps}
\eeq

\begin{table*}
 \begin{tabular}{lcccc}
\toprule
  Probe State& Particle Decomposition  &$g^{(2)}$&$g^{(2)}_{a,b}$&$\mathfrak{F}$ \\
\hline
$\hat S_a(\xi) \otimes \hat S_b(\xi) \ket{0,0}$ & $\sum_{n=0}^{\infty} \frac{\tanh^n \xi}{\cosh \xi} \left(\mathcal{\hat B} \ket{n,n}\right)$ & $3+1/\bar{n}$ & 1 & $\bar{n}^2+2 \bar{n}$\\
$\hat{\mathcal{B}}\ket{n,n}$ & - & $\frac{3}{2}-\frac{1}{\bar{n}}$ & $\frac{1}{2}-\frac{1}{\bar{n}}$ & $\frac{\bar{n}^2+2\bar{n}}{2}$ \\
$\frac{\ket{\alpha,0}+\ket{0,\alpha}}{\sqrt{2}}$ & $\sum_{n=0}^{\infty} e^{-\frac{|\alpha|^2}{2}}\frac{\alpha^n}{\sqrt{n!}} \left( \frac{\ket{n,0}+\ket{0,n}}{\sqrt{2}}\right)$ & 2 & 0 & $\bar{n}^2+\bar{n}$ \\
$\frac{\ket{n,0}+\ket{0,n}}{\sqrt{2}}$ & - & $2-\frac{2}{\bar{n}}$ & 0 & $\bar{n}^2$\\
$\hat S_a(\xi) \otimes \hat S_b(\xi) \mathcal{\hat B}\ket{1,0}$ & $\sum_{n=0}^{\infty} \frac{\sqrt{n+1}\tanh^n \xi}{\cosh^2 \xi} \left(\mathcal{\hat B}\ket{n+1,n} \right)$ & $\frac{9 \bar{n}^2+2 \bar{n}-11}{4 \bar{n}^2}$ & $\frac{3 \bar{n}^2-\bar{n}-1}{4 \bar{n}^2}$ & $\frac{3 \bar{n}^2+6 \bar{n}-5}{4}$ \\
$\mathcal{\hat B}\ket{n+1,n}$ & - & $\frac{3\bar{n}^2-2\bar{n}-1}{2 \bar{n}^2}$ & $\frac{(-1+\bar{n})^2}{2\bar{n}^2}$ & $\frac{\bar{n}(\bar{n}+2)-1}{2}$\\
$\left|\frac{\alpha}{\sqrt{2}},\frac{i \alpha}{\sqrt{2}}\right\rangle$ & $\sum_{n=0}^{\infty} e^{-\frac{|\alpha|^2}{2}}\frac{\alpha^n}{\sqrt{n!}} \left( \mathcal{\hat B}\ket{n,0} \right)$ & 1 & 1 & $\bar{n}$\\
$\mathcal{\hat B}\ket{n,0}$ & - &$1-\frac{1}{\bar n}$ & $1-\frac{1}{\bar n}$ & $\bar{n}$ \\
$\exp(\chi (\hat a^\dagger \hat b^\dagger- \hat a \hat b))\ket{0,0}$ & $\sum_{n=0}^{\infty} \frac{\tanh^n \chi}{\cosh \chi} \left(\ket{n,n} \right)$ & 4 & $4+\frac{2}{\bar{n}}$ & 0\\
$\ket{n,n}$ & -& $1-\frac{2}{\bar{n}}$ & 1 & 0\\
\toprule
 \end{tabular}
 \caption{The first order coherence function $g^{(2)}$, second order conference function $g^{(2)}_{a,b}$, and the quantum Fisher information $\mathfrak{F}$ as a function of the mean number of photons $\bar{n}$ are listed for various MZI path-symmetric probe states : Twin squeezed vacuum \cite{PhysRevLett.104.103602}, twin Fock states \cite{holland93}, entangled coherent state \cite{PhysRevLett.107.083601}, NOON state \cite{boto00}, amplified Bell state \cite{sahota13}, fraternal twin-Fock state \cite{lang14}, coherent states, and two-mode squeezed vacuum. The second column contains the decomposition of states with particle fluctuations in terms of states with fixed number of photons.  Note that the states with particle fluctuations and their fixed-particle-number projections have the same scaling of the QFI. The operator $\hat S_c(\xi)=\exp(i \xi \{ (\hat c^\dagger)^2+\hat c^2 \}/2)$ and $\mathcal{\hat B}=\exp(-i\frac{\pi}{2} \hat J_x)$ is the single-mode squeezer and beam splitter unitary respectively.
 \label{table1}}
\end{table*}

As discussed in Ref. \cite{hofman09}, path-symmetric states are ubiquitous in metrology, hence equation Eq. (\ref{ps}) holds for most probe states of interest (for example, see table \ref{table1}). Note that the first term in (\ref{Fmode}) and (\ref{ps}) is the classical (i.e. shot-noise) scaling term: it defines the optimal phase sensitivity attainable by a classical probe. In order to surpass the shot-noise limit, quantum correlations must be leveraged as described the remaining terms in (\ref{Fmode}) and (\ref{ps}) -- these Heisenberg limited terms add up to zero in the classical case of laser interferometry. We deduce that phase sensitivity is positively correlated with $g^{(2)}$ (intra-mode photon detection correlations) and negatively correlated with $g^{(2)}_{a,b}$ (inter-mode photon detection correlations). In addition, it is evident that mode entanglement is not vital for quantum metrology as Heisenberg scaling can be attained solely manipulating $g^{(2)}$ (e.g. consider the mode separable twin squeezed state probe, $\ket{\Psi}=\ket{S_a}\otimes\ket{S_b}$ \cite{sahota15}).

\section{The Particle Picture} \label{particlepic}

Now we will partition the Hilbert space of the probe state with respect to individual photons, which have two orthogonal internal state $\ket{\mu^{(i)}}$ and $\ket{\nu^{(i)}}$, where $i$ is a pseudo-label for the photons as discussed in section \ref{particleEnt}. Note that this is only possible in the case where the state in question has a well defined number of photons (\emph{i.e.} is an Eigenstate of the operator $\hat J_0$). When the state has particle fluctuations one can still talk about the particle entanglement contained in its $n$ particle sectors. In Table (\ref{table1}) we list several states previously used in Metrology with particle fluctuations and provide their decomposition in terms of states containing exactly $n$ particles.

 We can define the Pauli operators for the $i^{\text{th}}$ photon in the $\{\ket{\mu^{(i)}},\ket{\nu^{(i)}}\}$ basis:
\begin{align} 
\hat{\sigma}^{(i)}_{x} &=\ket{\mu^{(i)}}\bra{\nu^{(i)}}+\ket{\nu^{(i)}}\bra{\mu^{(i)}}, \label{x} \\
\hat{\sigma}^{(i)}_{y} &=-i \ket{\mu^{(i)}}\bra{\nu^{(i)}}+ i\ket{\nu^{(i)}}\bra{\mu^{(i)}}, \label{y} \\
\hat{\sigma}^{(i)}_{z} &= \ket{\mu^{(i)}}\bra{\mu^{(i)}}-\ket{\nu^{(i)}}\bra{\nu^{(i)}}. \label{z} 
\end{align}  
Let us denote $\hat{\boldsymbol\sigma}^{(i)} = (\hat{\sigma}^{(i)}_{x} , \hat{\sigma}^{(i)}_{y} , \hat{\sigma}^{(i)}_{z})$ and $\hat{\mathbf{J}} = \frac{1}{2} \sum_{i} \hat{\boldsymbol\sigma}^{(i)}$, where the sum is over all the particles in the state. Now we can define the collective spin operators
\beq
\hat{J}_{v}=\boldsymbol{v} \cdot \hat{\mathbf{J}}, 
\eeq
where the unit vector $\boldsymbol{v} \in \mathbb{R}^3$ specifies the spin direction. For directions $\boldsymbol{x} =(1,0,0), \boldsymbol{y}=(0,1,0),$ and $\boldsymbol{z}=(0,0,1)$, we obtain the collective spin operators
\begin{subequations}
\begin{align} 
\hat{J}_{x} &=\boldsymbol{x} \cdot \hat{\mathbf{J}}, \label{Jx} , \\ 
\hat{J}_{y} &=\boldsymbol{y} \cdot \hat{\mathbf{J}}, \label{Jy}, \\
\hat{J}_{z} &=\boldsymbol{z} \cdot \hat{\mathbf{J}}, \label{Jz} ,
\end{align} 
\end{subequations}   
which satisfy the SU(2) Lie algebra commutation relations \ref{su2}. 
Note that the Jordan-Schwinger isomorphism dictates the equivalence of (\ref{Schwinger1}), (\ref{Schwinger2}), (\ref{Schwinger3}) with (\ref{Jx}), (\ref{Jy}), (\ref{Jz}) . 

Now we can succinctly express the various MZI operations in terms of these operators; i.e. first BS: $\exp{(-i \frac{\pi}{2} \hat{J}_x)}$, phase-shift: $\exp{(-i \varphi \hat{J}_z)}$, final BS: $\exp{(i \frac{\pi}{2} \hat{J}_x)}$, and the complete MZI unitary: 
\begin{align} 
\hat{U}_{\text{MZI}} &= e^{-i \frac{\pi}{2} \, \hat{J}_x} e^{-i \varphi \hat{J}_z} \, e^{i \frac{\pi}{2} \hat{J}_x} \nonumber \\
&=e^{-i \varphi \hat{J}_y}.
\end{align}    
All unitary transformations of the form $\exp{(-i \gamma \, \boldsymbol{v} \cdot \hat{\mathbf{J}})}$ can be factorized for each particle as
\beq
e^{\-i \gamma \boldsymbol{v} \cdot \hat{\boldsymbol\sigma}^{(1)} } \otimes e^{-i \gamma \boldsymbol{v} \cdot \hat{\boldsymbol\sigma}^{(2)}} \otimes \dots \otimes e^{-i \gamma \boldsymbol{v} \cdot \hat{\boldsymbol\sigma}^{(n)}},
\eeq
where $\gamma$ is some arbitrary interaction constant. As noted in Ref.\cite{hyllus10}, this means that all MZI unitaries are local in the particle picture (this is not true in the mode picture). Since particle entanglement is preserved by MZI operations, we are enticed to investigate its role as a potential resources for  phase estimation. Hyllus \emph{et al.} have illustrated a positive correlation between a specific form of particle entanglement and phase sensitivity \cite{hyllus10, PhysRevA.85.022321}. 

\begin{table*}
 \begin{tabular}{l c c}
 \toprule
  \, & Mode Picture & Particle Picture \\
\hline
 Hilbert Space  &$\mathcal{H} = \mathcal{H}_a \otimes \mathcal{H}_b $&$\mathcal{H} = \mathcal{H}_1 \otimes \mathcal{H}_2 \otimes \dots \otimes \mathcal{H}_n$ \\
  Probe State & $\ket{\Psi}=e^{-i \frac{\pi}{2} \hat{J}_x}\ket{\Psi_o}$& $\ket{\Psi}=e^{-i \frac{\pi}{2} \hat{J}_x}\ket{\Psi_o}$ \\
$\hat{J}_x$&$\frac{1}{2}(\hat{a}^{\dagger} \hat{b} + \hat{b}^{\dagger} \hat{a} )$& $\frac{1}{2}\sum_{i=1}^{n} \hat{\sigma}_x^{(i)}$ \\
 $\hat{J}_y$& $- \frac{i}{2}( \hat{a}^{\dagger} \hat{b} - \hat{b}^{\dagger} \hat{a} )$ & $\frac{1}{2}\sum_{i=1}^{n} \hat{\sigma}_y^{(i)}$ \\
$\hat{J}_z$ &$\frac{1}{2} ( \hat{a}^{\dagger} \hat{a} - \hat{b}^{\dagger} \hat{b} )$& $\frac{1}{2}\sum_{i=1}^{n} \hat{\sigma}_z^{(i)}$ \\
QFI &$\bar{n} + \frac{\bar{n}^2}{2} (g^{(2)}-g^{(2)}_{a,b})$& $n \text{Var}[\hat{\sigma}_z^{(i)}] + n\left(n-1\right) \text{Cov}[\hat{\sigma}_z^{(i)}, \hat{\sigma}_z^{(j)}]$\\
\toprule
 \end{tabular}
 \caption{Comparison of the mode picture and particle picture}
 \label{particlemode}
\end{table*}

Now we are ready to analyze the QFI in the particle picture. Using expression (\ref{Jz}) for $\hat{J}_z$, we can write the QFI as follows.
\begin{align}
\mathfrak{F} &= 4 \text{Var}[ \hat{J}_z ] \\
&= \left \langle \left( \sum_{i=0}^{n} \hat{\sigma}_z^{(i)} \right)^2 \right \rangle - \left \langle \sum_{i=0}^{n} \hat{\sigma}_z^{(i)} \right \rangle^2
\end{align}
Expanding $( \sum_{i=0}^{n} \hat{\sigma}_z^{(i)})^2=  \sum_{i=0}^{n} \hat{\sigma}_z^{(i) 2} +  2 \sum_{i<j} \hat{\sigma}_z^{(i)} \otimes \hat{\sigma}_z^{(j)}$ and $\langle \sum_{i=0}^{n} \hat{\sigma}_z^{(i)} \rangle^2=\sum_{i=0}^{n} \langle \hat{\sigma}_z^{(i)}\rangle^2 + 2 \sum_{i<j} \langle \hat{\sigma}_z^{(i)}\rangle \langle \hat{\sigma}_z^{(j)} \rangle$, we obtain
\beq
\mathfrak{F} = \sum_{i=0}^{n} \text{Var}[\hat{\sigma}_z^{(i)}] +2\sum_{i<j} \left( \langle \hat{\sigma}_z^{(i)} \otimes \hat{\sigma}_z^{(j)} \rangle - \langle \hat{\sigma}_z^{(i)}\rangle \langle \hat{\sigma}_z^{(j)} \rangle\right), \nonumber
\eeq
where the first sum contains $n$ terms and the second terms contains $n(n-1)/2$ terms. Using the permutation symmetry of identical particles, we conclude $\langle \hat{\sigma}_z^{(i)} \rangle = \langle \hat{\sigma}_z^{(j)} \rangle$, $\langle \hat{\sigma}_z^{(i) 2} \rangle = \langle \hat{\sigma}_z^{(j) 2} \rangle$, and $ \langle \hat{\sigma}_z^{(i)} \otimes \hat{\sigma}_z^{(j)} \rangle =  \langle \hat{\sigma}_z^{(j)} \otimes \hat{\sigma}_z^{(i)} \rangle$ for all $i, j$. Therefore,  
\beq
\mathfrak{F} = n \text{Var}[\hat{\sigma}_z^{(i)}] + n\left(n-1\right) \text{Cov}[\hat{\sigma}_z^{(i)}, \hat{\sigma}_z^{(j)}], \label{partQFI}
\eeq
where $\text{Cov}[\hat{\sigma}_z^{(i)}, \hat{\sigma}_z^{(j)}]=\langle \hat{\sigma}_z^{(i)} \otimes \hat{\sigma}_z^{(j)} \rangle - \langle \hat{\sigma}_z^{(i)}\rangle \langle \hat{\sigma}_z^{(j)} \rangle$. Now we see that when there is no entanglement between particles, i.e. when $\text{Cov} [\hat{\sigma}_z^{(i)}, \hat{\sigma}_z^{(j)}]=0$, the maximum phase sensitivity is shot-noise limited, as evident from the remaining $n \text{Var}[\hat{\sigma}_z^{(i)}]$ term in (\ref{partQFI}). In order to surpass the SNL, the probe state must contain particle entanglement. The particle entanglement must be such that $\text{Cov} [\hat{\sigma}_z^{(i)}, \hat{\sigma}_z^{(j)}] > 0$. This result supports the findings of Ref. \cite{pezze09} and Ref. \cite{PhysRevLett.96.010401}: if the SNL is surpassed, then the probe state must be particle entangled.

\section{Conclusion}

The QFI for the MZI can be expressed in terms of first and second order Glauber coherence functions, as in Eq.(\ref{Fmode}) and Eq.(\ref{ps}). It is evident from these equations that the field intensity of the probe state (i.e. mean photon number) and its detection correlations (both inter-mode and intra-mode) are a resource of quantum-enhanced phase estimation. More specifically, the QFI is positively correlated with field intensity and the intra-mode coherence functions (\ref{ga}) and (\ref{gb}). Whereas, a negative correlation is exhibited with respect to the inter-mode coherence function (\ref{gab}). These resource parameters may be controlled experimentally in order to maximize phase sensitivity.  

Using the first quantization formalism, the QFI can also be expressed in terms of Pauli operators of the individual photons in the probe state (\ref{partQFI}). It follows from this form of the QFI that photon correlations, resulting from particle entanglement, are required to surpass the SNL. We note that for states with a fixed photon number, Eq.(\ref{Fmode}) and Eq.(\ref{partQFI}) are equivalent; this can be confirmed mathematically using the Jordan-Schwinger isomorphism.  

In studying quantum entanglement using the second quantization formalism of quantum mechanics, we conclude that mode entanglement is not required for quantum-enhanced phase estimation. On the other hand, we used the first quantization formalism to deduce the necessity of particle entanglement in order to obtain an advantage over classical phase estimation.

\bibliographystyle{osajnl}
\bibliography{new}

\end{document}